\documentclass[jkps,preprint,fleqn]{revtex4} 
\usepackage{graphicx}
\usepackage{amssymb}
\usepackage{amsmath}
\usepackage{bm}

\usepackage{kotex}

\usepackage{pifont}
\newcommand{\cmark}{\ding{51}} 
\newcommand{\xmark}{\ding{55}} 
\usepackage{booktabs}       
\usepackage{array}          

\pdfoutput=1

\usepackage{float} \usepackage{graphicx}
\usepackage{epstopdf}
\usepackage{graphics}
\usepackage{caption}
\usepackage{subfigure}
\usepackage{epsfig}
\usepackage{color}
\usepackage{multirow,array}

\usepackage{tensor}

\usepackage{adjustbox}  

\usepackage{tabularray}

\usepackage{hyperref}
\hypersetup{
	colorlinks=true,
	linkcolor=blue,
	filecolor=magneta,      
	urlcolor=blue,
}

\usepackage{tikz}
\usetikzlibrary{3d}

\newcommand{\tc}{\tilde{c}}

\newcommand{\tG}{\tilde{G}}

\newcommand{\GR}{\text{GR}}

\newcommand{\tkapp}{\tilde{\kappa}}

{\begin{description} \item[\emph{Remark :}]\em}%
{\end{description}}

{\begin{description} \item[\emph{Example #1 :}]\em}%
{\end{description}}

\makeindex

\begin{document} 
\setcounter{page}{0}

\title[]{Revisiting Varying Speed of Light in Cosmology: Insights from the Friedmann-Lemaître-Robertson-Walker Metric}
\author{Seokcheon \surname{Lee}}
\email{skylee@skku.edu}
\affiliation{Department of Physics, Institute of Basic Science, Sungkyunkwan University, Suwon 16419, Korea}

\date[]{Received }

\begin{abstract}
In the Friedmann-Lemaître-Robertson-Walker metric, a varying speed of light (VSL) reflects a change in the clock rate across hypersurfaces, described by the lapse function. This variation is not a dynamical field evolution but a consequence of coordinate choice, as the cosmic time coincides with the proper time of comoving observers due to the Weyl postulate. From an action principle including $\tc$, we derive that $\tc$ does not have its dynamics but imposes a constraint on the scale factor $a(t)$, indicating that it is not an independent degree of freedom. This insight reframes the VSL concept as a manifestation of gauge freedom in general relativity, wherein physical laws remain invariant under smooth coordinate transformations. Here, gauge refers to the freedom of choosing the temporal coordinate (\textit{e.g.}, setting the lapse $N(t) \neq 1$), which determines how the speed of light appears in the cosmological equations. Recognizing $\tc$ as a coordinate-dependent quantity offers a new interpretation of cosmological time and observational tensions, such as the Hubble tension, without invoking new physical fields. This redefinition opens a novel theoretical pathway in interpreting cosmic expansion within a consistent relativistic framework.
\end{abstract}



\maketitle

\tableofcontents

\section{Introduction}
\label{sec:intro}

The minimally extended varying speed of light (meVSL) model is formulated within the framework of the Friedmann-Lemaître-Robertson-Walker (FLRW) metric, which adheres to the cosmological principle (CP) by ensuring that the universe remains spatially homogeneous and isotropic on large scales \cite{Lee:2020zts,Lee:2023bjz,Lee:2024mal}. This principle requires that physical laws governing the universe avoid introducing preferred directions or locations. A crucial aspect of maintaining these symmetries is the conservation of adiabaticity, as any net energy flux within the cosmic medium would establish a privileged frame of reference, thereby violating isotropy \cite{Lee:2022heb}.

In the context of the meVSL model, the variation of the speed of light over cosmic time must be accompanied by corresponding changes in other fundamental physical constants to ensure the internal consistency of the theoretical framework. In particular, the Planck constant must evolve consistently with the varying speed of light to preserve the fundamental relationships governing quantum mechanics and thermodynamics to explain the adiabatic expansion \cite{Lee:2022heb}. More broadly, the cosmological evolution of additional physical quantities, including those governing electromagnetic interactions and relativistic dynamics, must be induced to maintain compatibility with all locally verified physical laws, such as special relativity and Maxwell’s equations \cite{Lee:2020zts,Lee:2023bjz,Lee:2024mal,Lee:2022heb}.

Unlike conventional VSL theories, which typically invoke explicit mechanisms to drive changes in $\tc$ as a function of the cosmic scale factor $a(t)$, the meVSL model does not require such an ad hoc prescription. Instead, it introduces a generalized condition on cosmological time dilation (CTD), leading to a time-dependent speed of light as a natural consequence. In the meVSL model, the lapse function determines the difference in clock rates across hypersurfaces, reflecting the time-dependent variation in the speed of light \cite{Lee:2024zcu}.  Rather than modifying the FLRW metric itself, the meVSL framework allows the variation of $\tc$ to emerge dynamically from the underlying structure of cosmological time without imposing a specific functional form $\tc(a)$.  

This distinguishing feature sets meVSL apart from traditional VSL models, which often assume or necessitate an explicit mechanism driving the evolution of the speed of light \cite{Avelino:1999is,Belinchon:1999kq,Avelino:2000ph,Szydlowski:2002kz,Magueijo:2003gj,Shojaie:2004sq,Shojaie:2004xw,Balcerzak:2013kha,Balcerzak:2014rga,Franzmann:2017nsc,Hanimeli:2019wrt,Skara:2019usd,Bhattacharjee:2020fgl,Gupta:2020anq,Cuzinatto:2022mfe,Cuzinatto:2022vvy,Cuzinatto:2022dta,Bileska:2024odt,Coleman:1997xq,Albrecht:1998ir,Barrow:1998df,Barrow:1999is,Bassett:2000wj,Jacobson:2000xp,Magueijo:2000zt,Clayton:1998hv,Drummond:1999ut,Clayton:1999zs,Liberati:2000us,Clayton:2000xt,Drummond:2001rj,Amelino-Camelia:1996bln,Amelino-Camelia:1997ieq,Ellis:1999sd,Amelino-Camelia:2000bxx,Amelino-Camelia:2000cpa,Ellis:2000sf,Kowalski-Glikman:2001vvk,Bruno:2001mw,Magueijo:2001cr,Amelino-Camelia:2002uql,Magueijo:2002pg,Moffat:1992ud,Manida:1999rx,Barrow:1999st,Stepanov:1999ax,Magueijo:2000au,Moffat:2002nm,Kaelbermann:1998hu,Randall:1999ee,Randall:1999vf,Kiritsis:1999tx,Chung:1999xg,Alexander:1999cb,Ishihara:2000nf,Csaki:2000dm,Youm:2001sw,Youm:2001zk,Grojean:2001pv,Youm:2001zp,Drummond:1979pp,Novello:1988ma,Barton:1989dq,Scharnhorst:1990sr,Shore:1995fz,Colladay:1995qb,Coleman:1998ti,Bertolami:1999da,Shore:2000bs,Greenberg:2002uu,Teyssandier:2003qh,Shore:2003zc,Blasone:2003wf,Alexander:2001dr,Burgess:2002tb}.  By embedding the variation of fundamental constants within the broader context of cosmological expansion and time dilation, the meVSL model provides a more generalized and self-consistent approach to exploring the implications of a non-fixed speed of light in the expanding universe.

 The outline of this manuscript is as follows. In Section 2, we provide a brief overview of the existing VSL models.
Section 3 introduces the FLRW metric in the context of the meVSL model, which serves as the foundation for our analysis. Section 4 discusses the Einstein-Hilbert (EH) action, providing the theoretical framework for deriving the field equations. In section 5, we derive the Einstein field equations (EFEs), including the Ricci and Einstein tensors, the energy-momentum tensor, and the Friedmann equations. Section 6 is dedicated to the equation of motion for the speed of light, exploring its implications within the meVSL framework. Finally, we conclude with a discussion of our findings and their broader implications in Section 6.

\section{A Brief Summary of Existing VSL Theories}

\begin{table}[htbp]
\centering
\caption{Comparison of representative VSL theories and the meVSL framework.}
\label{tab:VSLComparison}
\renewcommand{\arraystretch}{1.35} 

\resizebox{\textwidth}{!}{%
 \begin{tabular}{
    >{\raggedright\arraybackslash}p{3.2cm}  
    >{\raggedright\arraybackslash}p{4.5cm}  
    >{\centering\arraybackslash}p{1.2cm}    
    >{\centering\arraybackslash}p{1.5cm}    
    >{\centering\arraybackslash}p{1.5cm}    
    >{\centering\arraybackslash}p{1.6cm}    
    >{\raggedright\arraybackslash}p{2.8cm}  
    >{\raggedright\arraybackslash}p{2.8cm}  
}
\toprule
\hline
\textbf{Model} & \textbf{Variation Mechanism} & \textbf{Dyn} & \textbf{LI} & \textbf{New Fields} & \textbf{GR Covariance} & \textbf{Obs. Target} & \textbf{References} \\
\midrule
\hline
Hard Lorentz breaking & Preferred frame with $c(t)$ imposed explicitly & \xmark/\cmark & \xmark & \xmark/\cmark & \xmark & Conceptual variation & \cite{Coleman:1997xq,Albrecht:1998ir,Barrow:1998df,Barrow:1999is,Bassett:2000wj,Jacobson:2000xp,Magueijo:2000zt} \\
Bimetric VSL & $\hat{g}_{\mu\nu} = g_{\mu\nu} + B\partial_\mu\phi\partial_\nu\phi$ & \cmark & Partial & \cmark & Partial & Propagation delay & \cite{Clayton:1998hv,Drummond:1999ut,Clayton:1999zs,Liberati:2000us,Clayton:2000xt,Drummond:2001rj}  \\
Color-dependent VSL & $c(\nu)$ varies by frequency due to vacuum dispersion & \xmark & \xmark & \xmark & \xmark & High-energy astrophysics & \cite{Amelino-Camelia:1996bln,Amelino-Camelia:1997ieq,Ellis:1999sd,Amelino-Camelia:2000bxx,Amelino-Camelia:2000cpa,Ellis:2000sf,Kowalski-Glikman:2001vvk,Bruno:2001mw,Magueijo:2001cr,Amelino-Camelia:2002uql,Magueijo:2002pg} \\
Lorentz-invariant VSL & $c(x^\mu)$ as a scalar field in covariant framework & \cmark & \cmark & \cmark & \xmark & Model-dependent & \cite{Moffat:1992ud,Manida:1999rx,Barrow:1999st,Stepanov:1999ax,Magueijo:2000au,Moffat:2002nm} \\
String/M-theory & $c$ varies via compactification or brane motion & \cmark & \cmark & \cmark & Partial & Early-universe physics & \cite{Kaelbermann:1998hu,Randall:1999ee,Randall:1999vf,Kiritsis:1999tx,Chung:1999xg,Alexander:1999cb,Ishihara:2000nf,Csaki:2000dm,Youm:2001sw,Youm:2001zk,Grojean:2001pv,Youm:2001zp} \\
Field-theory VSL & $c = c(\phi)$ via scalar field $\phi$ & \cmark & Model-dep. & \cmark & \xmark & CMB, LSS, BBN &  \cite{Drummond:1979pp,Novello:1988ma,Barton:1989dq,Scharnhorst:1990sr,Shore:1995fz,Colladay:1995qb,Coleman:1998ti,Bertolami:1999da,Shore:2000bs,Greenberg:2002uu,Teyssandier:2003qh,Shore:2003zc,Blasone:2003wf} \\
Hybrid models & Combine metric and scalar field frameworks & \cmark & Model-dep. & \cmark & Model-dep. & Mixed datasets & \cite{Alexander:2001dr,Burgess:2002tb} \\
meVSL & Lapse function $N(t)$ chosen via coordinate freedom $\Rightarrow \tc(t)$ & \xmark & \cmark & \xmark & \cmark & CTD, $H_0$ tension & \cite{Lee:2020zts,Lee:2023ucu,Lee:2024kxa,Lee:2021ona,Lee:2023rqv,Lee:2024nya} \\
\bottomrule
\hline
\end{tabular}}
\end{table}

Over the past decades, various frameworks for implementing a varying speed of light (VSL) have been proposed. These can be categorized as follows.

\begin{itemize}
    \item \textbf{Hard breaking of Lorentz symmetry}:  
    These models explicitly violate Lorentz invariance by introducing a preferred frame or absolute structure in spacetime \cite{Coleman:1997xq,Albrecht:1998ir,Barrow:1998df,Barrow:1999is,Bassett:2000wj,Jacobson:2000xp,Magueijo:2000zt}.  While allowing for a straightforward variation of $c$, they often face serious theoretical difficulties, such as conflicts with the principle of relativity and difficulties in formulating a consistent quantum field theory.
    \item \textbf{Bimetric VSL theories}:  
    In these models, two distinct metrics are introduced: one for gravitational phenomena and another for photon propagation  \cite{Clayton:1998hv,Drummond:1999ut,Clayton:1999zs,Liberati:2000us,Clayton:2000xt,Drummond:2001rj}  
\begin{align}
\hat{g}_{\mu\nu} = g_{\mu\nu} + B \partial_{\mu} \phi \partial_{\nu} \phi \label{bimetric} \,,
\end{align}
where $g_{\mu\nu}$ is a graviton metric, $\hat{g}_{\mu\nu}$ is a matter metric, and $\phi$ is a scalar field. Although they preserve some gravitational structures, bimetric theories typically introduce additional degrees of freedom, complicating their dynamics and raising issues of stability and causality.
    \item \textbf{Color-dependent speed of light}:  
    These frameworks allow the speed of light to vary depending on photon frequency, leading to a vacuum dispersion relation \cite{Amelino-Camelia:1996bln,Amelino-Camelia:1997ieq,Ellis:1999sd,Amelino-Camelia:2000bxx,Amelino-Camelia:2000cpa,Ellis:2000sf,Kowalski-Glikman:2001vvk,Bruno:2001mw,Magueijo:2001cr,Amelino-Camelia:2002uql,Magueijo:2002pg}.  
    However, such color-dependent variations are tightly constrained by astrophysical observations (e.g., gamma-ray bursts and gravitational wave counterparts), limiting their viability.
    \item \textbf{Lorentz invariant VSL theories}:  
    It is possible to construct VSL theories that preserve the essence of Lorentz invariance. One approach, proposed by Moffat~\cite{Moffat:1992ud}, involves spontaneous breaking of local Lorentz symmetry, where a Lorentz scalar field acquires a time-like vacuum expectation value, selecting a preferred frame and breaking $O(3,1)$ symmetry down to $O(3)$. In this framework, the speed of light undergoes a phase transition to its present small value, and the spontaneous breaking direction provides an explanation for the arrow of time and the second law of thermodynamics. Another approach~\cite{Magueijo:2000zt} defines a covariant and locally Lorentz invariant theory in which the speed of light varies through a scalar field $\psi = \log(c/c_0)$. Depending on model parameters, these theories can either be equivalent to scalar-tensor theories or define genuinely new frameworks. In such models, the cosmological constant $\Lambda$ can vary with $c$, acting as a potential for $\psi$, typically scaling as $\Lambda \propto (c/c_0)^n$ \cite{Manida:1999rx,Barrow:1999st,Stepanov:1999ax,Magueijo:2000au,Moffat:2002nm}.
  \item \textbf{String/M-theory efforts}:  
    Within string or M-theory, variations in $c$ can arise naturally from dynamics of extra dimensions or evolving scalar fields (moduli) \cite{Kaelbermann:1998hu,Randall:1999ee,Randall:1999vf,Kiritsis:1999tx,Chung:1999xg,Alexander:1999cb,Ishihara:2000nf,Csaki:2000dm,Youm:2001sw,Youm:2001zk,Grojean:2001pv,Youm:2001zp}.  
    Although these scenarios are theoretically appealing, they heavily depend on assumptions about the compactification scheme and stabilization mechanisms, many of which remain speculative.
   \item \textbf{Field theory VSL predictions}:  
    In these approaches, $c$ is treated as an effective coupling that varies due to a dynamical scalar field \cite{Drummond:1979pp,Novello:1988ma,Barton:1989dq,Scharnhorst:1990sr,Shore:1995fz,Colladay:1995qb,Coleman:1998ti,Bertolami:1999da,Shore:2000bs,Greenberg:2002uu,Teyssandier:2003qh,Shore:2003zc,Blasone:2003wf}.  
    While more closely tied to standard field theory techniques, ensuring gauge invariance and compatibility with the standard model requires intricate constructions.
    \item \textbf{Hybrids}:  
    Some models combine the above elements, for instance employing a bimetric structure alongside a dynamical scalar field  \cite{Alexander:2001dr,Burgess:2002tb}.  
    These hybrids aim to balance strengths but often inherit the theoretical challenges of multiple categories simultaneously.
\end{itemize}

Each of these models faces nontrivial obstacles, especially in preserving fundamental symmetries and aligning with observational data.  
In contrast, the meVSL model introduces the variation of $\tc$ via the lapse function, interpreting changes in $\tc$ as coordinate effects rather than introducing a new physical degree of freedom.  
This allows meVSL to preserve the CP and the principle of relativity at each moment, offering a conceptually simpler and more symmetric framework compared to traditional VSL models. The differences between the previously discussed VSL models and the meVSL model are summarized in Table~\ref{tab:VSLComparison}.  Here, \textbf{Dyn} refers to whether the speed of light is treated as a dynamical quantity governed by field equations, and \textbf{LI} indicates Lorentz invariance. \textbf{Obs. Target} denotes the primary observational phenomena each model aims to address.

\section{Friedmann-Lemaître-Robertson-Walker metric}
\label{sec:RWm}

\begin{figure}
	\begin{center}
	\includegraphics[width=0.9\textwidth]{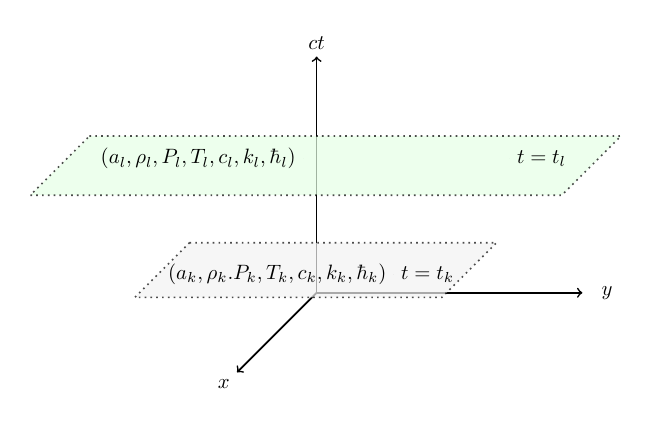} 
	\caption{At $t = t_k$, the values of physical quantities and constants, such as $a_k$, $\rho_k$, $P_k$,  $T_k$, $\tc_k$,  $k_{k}$, and $\hbar_k$, are fixed and independent of spatial position on the $t=t_k$ hypersurface. As the universe expands, these quantities and constants transition to $a_l$, $\rho_l$, $P_l$,  $T_l$, $\tc_l$,  $k_l$, and $\hbar_l$. The CP and Weyl’s postulate do not restrict $c_k$ to be equal to $c_l$.  Its value is determined by the CTD relation. }
	\label{Fig1}
	\end{center}
\end{figure}

The FLRW metric is based on the CP, which posits that the universe is homogeneous and isotropic at every moment. Mathematically, this implies that the Lie derivatives of the Killing vectors (KVs) responsible for generating spatial isotropy and homogeneity must vanish, ensuring the metric remains invariant under both translations and rotations \cite{Lee:2024mal,Ryder09}. The general metric for an isotropic and homogeneous space at a given time $t_l'$ can be expressed as
\begin{equation}
g_{\mu\nu}^{(\text{CP})}(t_l') = \text{diag} \left( g_{00}(t_l') \,,  \frac{A(t_l')}{1-Kr^2} \,,  A(t_l') r^2 \,,  A(t_l') r^2 \sin^2 \theta \right) \,.
\end{equation}
The corresponding 4-dimensional line element is given by
\begin{equation}
ds^2(t_l') = c_l^2 g_{00}(t_l') dt_l'^2 + A(t_l') \left[ \frac{dr^2}{1-Kr^2} + r^2 d \theta^2 + r^2 \sin^2 \theta d \phi^2 \right] \label{dstlp} \,,
\end{equation}
where $A(t_l') (= a^2(t_l'))$ represents the square of the scale factor and \( g_{00}(t_l') \) is explicitly given by \(- g_{00}(t_l') \equiv N^2(t_l') \), where \( N(t_l') \) is the lapse function \cite{Lee:2024zcu,Ryder09}. By adopting Weyl's postulate to extend the metric described in Eq.~\eqref{dstlp} to cosmic time $t$, we can represent the line element as
\begin{align}
ds^2 = - \tc(t)^2 dt^2 + a(t)^2 \left[ \frac{dr^2}{1-Kr^2} + r^2  \left( d \theta^2 + \sin^2 \theta d \phi^2 \right)  \right] \equiv - \tc(t)^2 dt^2 + a(t)^2 dl_{3\textrm{D}}^2 \label{dstgen} \,,
\end{align} 
where the speed of light is treated as a function of time, differing from its conventional role in the standard FLRW metric (\textit{i.e.}, $N = 1$). At first glance, this formulation might appear unconventional or even incorrect. However, as shown in Figure~\ref{Fig1}, the standard FLRW metric suggests that on hypersurfaces where $t_l$ or $t_k$ are constant, various physical quantities—such as the scale factor $a_l = a(t_l)$, mass density $\rho_l$, pressure $P_l$, temperature $T_l$, speed of light $\tc_l$, Boltzmann constant $k_l$, and Planck constant $\hbar_l $—remain constant across all spatial positions in three-dimensional space. However, following Weyl’s postulate, these quantities can evolve as functions of cosmic time $t$, reflecting cosmological redshift effects, as illustrated in Figure \ref{Fig1}.

In traditional cosmological models, physical constants, including the speed of light, are typically assumed to be invariant over time. However, this assumption—that $\tc_l$ remains equal to $\tc_k$ across cosmic time—is not an intrinsic requirement of the two fundamental conditions used to derive the FLRW metric, namely the CP and Weyl’s postulate. Instead, the constancy of the speed of light is related to CTD. It is important to note that General Relativity (GR) does not impose any fundamental law dictating that the speed of light must remain unchanged. As the universe evolves from $t_k$ to $t_l$, physical parameters such as $a(t)$, $\rho(t)$, $P(t)$, and $T(t)$ evolve as functions of time. Their precise behavior is determined by solving EFEs and Bianchi’s identity (BI), taking into account the equation of the state (e.o.s) of the cosmic fluid.

In cosmology, observable physical quantities are typically expressed not as functions of the cosmic time $t$, but rather as functions of the cosmological redshift $z$ or equivalently the scale factor $a(t) = 1/(1+z)$. For instance, quantities such as wavelength and temperature are redshifted due to the expansion of the universe, which reflects the geometric properties of spacetime. From this perspective, if dimensionful observables naturally evolve with redshift, it becomes reasonable to consider the possibility that certain physical constants, particularly the speed of light $\tc$, might also be expressed as functions of redshift. 

In the FLRW metric, the time coordinate is given by $x^0 = ct$, suggesting that if the lapse function is not fixed, the combination $\tc(t)$ could be interpreted as a function rather than a constant as shown in Eq.~\eqref{NmeVSL}. This opens up the theoretical possibility of extending standard cosmological models to include a varying speed of light without introducing new dynamical fields, but rather through the gauge freedom inherent in GR \cite{Lee:2024zcu}. Such an interpretation provides a framework in which the effects traditionally attributed to a constant $\tc$ might instead arise from the coordinate choice and could offer new insights into cosmological phenomena.

The derivation of redshift follows from the geodesic equation for a propagating light wave, where the condition $ds^2 = 0$ holds, as given in Eq.~\eqref{dstgen}. The spatial separation $dl_{3\textrm{D}}$ remains consistent over time in comoving coordinates  \cite{Lee:2024mal,Lee:2024zcu}. Building upon this framework, the expression for radial light propagation is given by
\begin{align}
d l_{3\textrm{D}} &= \frac{c(t_i) dt_i}{a(t_i)} \quad : \quad \frac{\tc_1 dt_1}{a_1} = \frac{\tc_2 dt_2}{a_2} \Rightarrow \begin{cases} \tc_1 = \tc_2 = c & \textrm{if} \quad \frac{dt_1}{a_1} = \frac{dt_2}{a_2} \qquad \textrm{SMC} \\ 
\tc_1 = \frac{f(a_2)}{f(a_1)} \frac{a_1}{a_2} \tc_2 & \textrm{if} \quad \frac{dt_1}{f(a_1)} = \frac{dt_2}{f(a_2)} \quad \textrm{VSL} \\ \tc_1 = \left( \frac{a_1}{a_2}\right)^{\frac{b}{4}} \tc_2 & \textrm{if} \quad \frac{dt_1}{a_1^{1-\frac{b}{4}}} = \frac{dt_2}{a_2^{1-\frac{b}{4}}} \quad \textrm{meVSL}  \end{cases} \,, \label{dl3D}
\end{align}
where $dt_i = 1/\nu(t_i)$ represents the time interval between successive wave crests at $t_i$, corresponding to the inverse of the frequency $\nu_i$ at that moment, and $f(a_i)$ is an arbitrary function of the scale factor $a(t_i)$.

In the Standard Model of Cosmology (SMC), an additional assumption is imposed: the speed of light remains constant at $c$. This assumption arises from the framework of GR, where $c$ is traditionally treated as a fundamental constant. As a result, the relationship between CTD and the inverse of the scale factor $a(t)$ at different time slices, $t_1$ and $t_2$, follows directly. However, this relationship is not derived from a fundamental physical law. In contrast, if the speed of light varies over time, as proposed in the meVSL model, the conventional redshift relation would require modification.

 In an expanding universe, transitioning from one hypersurface to another leads to an increase in the scale factor, which naturally causes the cosmological redshift of various physical quantities such as mass density and temperature. However, determining CTD cannot be conclusively made based solely on the CP and Weyl’s postulate within the framework of the FLRW metric. Instead, confirming such relationships requires experimental data. Direct observations of supernovae (SNe) light curves and spectra, aimed at assessing the decay times of distances, have contributed to efforts to measure CTD \cite{Leibundgut:1996qm,SupernovaSearchTeam:1997gem,Foley:2005qu,Blondin:2007ua,Blondin:2008mz,Lee:2023ucu,DES:2024vgg,Lee:2024kxa}. Another research avenue explores CTD by analyzing the elongation of peak-to-peak timescales in gamma-ray bursts (GRBs) \cite{Norris:1993hda,Wijers:1994qf,Band:1994ee,Meszaros:1995gj,Lee:1996zu,Chang:2001fy,Crawford:2009be,Zhang:2013yna,Singh:2021jgr}.  In addition, the light curves of distant quasars (QSOs) have been examined for potential CTD effects \cite{Hawkins:2001be,Dai:2012wp,Lewis:2023jab}. Despite these efforts, current observational data has not yet confirmed a clear relationship between CTD and the predictions made by the SMC. Additionally, the FLRW model lacks a defined mechanism for conclusively determining CTD. Consequently, exploring the potential of VSL within these observations remains valuable, provided the results align with predictions from the SMC.
Given the theoretical absence of CTD in the FLRW model, considering this relationship as a function $f(a)$ of the scale factor allows the speed of light to be expressed as 
\begin{align} \tc(t_1) = \frac{f(a_2)}{f(a_1)} \frac{a(t_1)}{a(t_2)} \tc(t_2) \label{cVSL}. \end{align} 
This highlights that, while the general application of the VSL model within GR cannot be conclusively asserted, it emerges as a natural consequence of an expanding universe as described by the FLRW metric. The meVSL model represents a specific case of VSL, defined by $f(a) = a^{1-b/4}$ \cite{Lee:2020zts,Lee:2023bjz}.

\section{Einstein-Hilbert action}\label{sec:HEaction}

We use the EH action to derive the EFEs via the principle of least action. In the VSL model, the speed of light is a function of cosmic time. However, if the speed of light is the only varying quantity, the Palatini identity term introduces an issue in recovering the EFEs. To resolve this, we must allow the gravitational constant ($\tG$) to vary with cosmic time as well, ensuring the correct form of the EFEs can be obtained from the EH action \cite{Lee:2020zts}. This condition ensures that the combination of the gravitational constant and speed of light in the EH action, given by the so-called the Einstein gravitational constan $\tkapp \equiv 8 \pi \tG/\tc^4$, remains independent of cosmic time.
The EH action in the VSL model is written as 
\begin{align} S &\equiv \int \Biggl[ \frac{1}{2 \tkapp} \left( R - 2 \Lambda \right) + \mathcal{L}_i \Biggr] \sqrt{-g} dt d^3x \label{SHmp} \,, \end{align} 
where $\mathcal{L}_i = \rho_i (1 + \omega_i) \tc^2$ represents the Lagrangian density for perfect fluids (\textit{i.e.}, matter and radiation). As we will demonstrate, both the speed of light and the gravitational constant must evolve cosmologically for the EFEs of GR to be recovered from the EH action in the meVSL framework.
The variation of the action with respect to the inverse metric must vanish, which leads to the following 
\begin{align} \delta S &= \int \left( \left[ \frac{\left( R - 2 \Lambda \right)}{2 \tkapp} \right] \delta \left( \sqrt{-g} \right)  + \frac{1}{2\tkapp} \sqrt{-g} \delta R \right) dt d^3 x + \int \delta \left( \sqrt{-g} \mathcal{L}{m} \right) dt d^3 x \nonumber \\ &= \int \frac{\sqrt{-g}}{2 \tkapp} \left[ R{\mu\nu} - \frac{1}{2} g_{\mu\nu} \left( R - 2 \Lambda \right) - \tkapp T_{\mu\nu} \right] \delta g^{\mu\nu} dtd^3 x \nonumber \\ &+ \int \frac{\sqrt{-g}}{2 \tkapp} \left[ \nabla_{\mu} \nabla_{\nu} - g_{\mu\nu} \Box \right] \delta g^{\mu\nu} dtd^3 x \label{deltaSHmp} \,. \end{align} 
To preserve the EFEs, the second term (the Palatini identity term) in Eq.~\eqref{deltaSHmp} must vanish. This implies that $\tkapp$ must remain constant even though both $\tc$ and $\tG$ evolve with cosmic time 
\begin{align}
&\tkapp = \text{const} \quad \Rightarrow \quad \frac{\tG_0}{\tc_0^4} = \frac{\tG}{\tc^4} \quad \textrm{if} \quad \tc = \tc_0 f(a) \quad , \quad \tG = \tG_{0} f(a)^4 \label{tkappaconstmp} \,,
\end{align}
where we set $a_0 = 1$, and $\tc_0$ and $\tG_0$ represent the present values of the speed of light and gravitational constant, respectively. Using these relationships, we obtain the EFEs, including the cosmological constant 
\begin{align} &R_{\mu\nu} - \frac{1}{2} g_{\mu\nu} R + \Lambda g_{\mu\nu} \equiv G_{\mu\nu} + \Lambda g_{\mu\nu}  = \frac{8 \pi \tG}{\tc^4} T_{\mu\nu} \label{tEFEmp} \,, \end{align} 
which maintains the standard covariant form of the EFEs in GR. This demonstrates that a VSL can be accommodated within the covariant framework of GR, without violating its fundamental symmetry under general coordinate transformations.

\section{Einstein Field Equations} \label{sec:EFEs}

We need to derive the EFEs using the FLRW metric and a perfect fluid within the framework of a VSL model. From these equations, we obtain the modified Friedmann equations, which can then be compared with those of the standard model to analyze the differences.

\subsection{Ricci and Einstein tensors}\label{subsec:RS}
We now derive the EFEs for the VSL model using the FLRW metric \cite{Lee:2020zts,Lee:2025rpw}. The Christoffel symbols $\Gamma^{\mu}_{\,\,\nu\lambda}$ for the FLRW metric in Eq.~\eqref{dstgen} are given by 
\begin{align} 
&\Gamma^{\mu}_{\,\,\nu\lambda} \equiv \frac{1}{2} g^{\mu\alpha} \left( g_{\alpha\nu,\lambda} + g_{\alpha\lambda,\nu} - g_{\nu\lambda,\alpha} \right) \label{Gamma}\,, \\ &\Gamma^{0}_{\,\,ij} = \frac{a\dot{a}}{\tc} \gamma{ij} \quad , \quad \Gamma^{i}_{\,\,0j} = \frac{1}{\tc}  \frac{\dot{a}}{a} \delta^i_j \quad , \quad \Gamma^{i}_{\,\,jk} = ^{s}\Gamma^{i}_{\,\,jk}  \label{Gammacomp} \,, \end{align} 
where $^{s}\Gamma^{i}_{\,\,jk}$ denotes the Christoffel symbols for the spatial part of the metric $\gamma_{ij}$. As seen in Eq.~\eqref{Gammacomp}, the form of the Christoffel symbols in the VSL model is the same as in GR, with the only distinction being that $\tc$ varies as a function of the scale factor.
The Riemann curvature tensor, which describes the curvature of the manifold, is given by 
\begin{align} 
&\tensor{R}{^\alpha_\beta_\mu_\nu} = \tensor{\Gamma}{^\alpha_\beta_\nu_,_\mu} - \tensor{\Gamma}{^\alpha_\beta_\mu_,_\nu} + \tensor{\Gamma}{^\alpha_\lambda_\mu} \tensor{\Gamma}{^\lambda_\beta_\nu} - \tensor{\Gamma}{^\alpha_\lambda_\nu} \tensor{\Gamma}{^\lambda_\beta_\mu} \label{Rabmu} \,, \\ &\tensor{R}{^0_i_0_j} = \frac{\gamma_{ij}}{\tc^2} \left( a \ddot{a} - \dot{a}^2 \frac{d \ln \tc}{d \ln a} \right) \quad , \quad \tensor{R}{^i_0_0_j} = \frac{\delta^{i}{j}}{\tc^2} \left( \frac{\ddot{a}}{a} - \frac{\dot{a}^2}{a^2} \frac{d \ln \tc}{ d \ln a}  \right) \,, \label{R0i0j} \\ &\tensor{R}{^i_j_k_m} = \frac{\dot{a}^2}{\tc^2} \left( \delta^{i}{k} \gamma_{jm} - \delta^i_m \gamma_{jk} \right) + \tensor[^s]{R}{^i_j_k_m} \quad , \quad \tensor[^s]{R}{^i_j_k_m} = k \left( \delta^i_k \gamma_{jm} - \delta^i_m \gamma_{jk} \right) \,. \label{Rijkm} \end{align} 
While the Christoffel symbols in the VSL model follow the same form as in GR, the Riemann curvature tensors differ. This difference arises because the Riemann curvature tensors are derived from the derivatives of the Christoffel symbols, which include the time-varying speed of light. This leads to correction terms, such as the factor $H^2 \frac{d \ln \tc}{d \ln a}$, appearing in both $\tensor{R}{^0_i_0_j}$ and $\tensor{R}{^i_0_0_j}$.
The Ricci curvature tensors, which measure the deformation of space as one moves along geodesics, are obtained by contracting the Riemann curvature tensors from Eqs.~\eqref{R0i0j} and \eqref{Rijkm} 
\begin{align} 
R_{\mu\nu} &= \tensor{\Gamma}{^\lambda_\mu_\nu_,_\lambda} - \tensor{\Gamma}{^\lambda_\mu_\lambda_,_\nu} + \tensor{\Gamma}{^\lambda_\mu_\nu} \tensor{\Gamma}{^\sigma_\lambda_\sigma} - \tensor{\Gamma}{^\sigma_\mu_\lambda} \tensor{\Gamma}{^\lambda_\nu_\sigma} \label{Rmn} \,, \\ R_{00} &= -\frac{3}{\tc^2} \left( \frac{\ddot{a}}{a} - \frac{\dot{a}^2}{a^2} \frac{d \ln \tc}{ d \ln a}  \right) \quad , \quad R_{ij} = \frac{\gamma_{ij}}{\tc^2} a^2 \left( 2 \frac{\dot{a}^2}{a^2} + \frac{\ddot{a}}{a} + 2 k \frac{\tc^2}{a^2} - \frac{\dot{a}^2}{a^2} \frac{d \ln \tc}{ d \ln a}  \right) \label{Rij} \,. \end{align} 
The time-dependent variation of $\tc$ introduces correction terms in both $R_{00}$ and $R_{ij}$.
Finally, the Ricci scalar can be obtained by taking the trace of the Ricci tensor 
\begin{align} R &= \frac{6}{\tc^2} \left( \frac{\ddot{a}}{a} + \frac{\dot{a}^2}{a^2} + k \frac{\tc^2}{a^2} - \frac{\dot{a}^2}{a^2} \frac{d \ln \tc}{ d \ln a}  \right) =  \frac{6}{\tc^2} \left( \frac{\ddot{a}}{a} + \frac{\dot{a}^2}{a^2} + k \frac{\tc^2}{a^2} - \frac{\dot{a}}{a} \frac{\dot{\tc}}{\tc}  \right)  \label{Rmp} \,, \end{align} 
where the time-dependent speed of the light effect is evident in the final term.

The Einstein tensor is defined as follows
\begin{align}
G_{\mu\nu} = R_{\mu\nu} - \frac{1}{2} g_{\mu\nu} R \label{Gmunu} \,.
\end{align}
Using the previously derived Ricci tensors~\eqref{Rij} and Ricci scalar~\eqref{Rmp}, we can now obtain the components of the Einstein tensor
\begin{align}
G_{00} = \frac{3}{\tc^2} \left[ \frac{\dot{a}^2}{a^2} + k \frac{\tc^2}{a^2} \right] \quad , \quad G_{ij} = -\frac{g_{ij}}{\tc^2} \left[ 2 \frac{\ddot{a}}{a} + \frac{\dot{a}^2}{a^2} + k \frac{\tc^2}{a^2} - 2 \frac{\dot{a}^2}{a^2} \frac{d \ln \tc}{d \ln a} \right] \,. \label{G00Gij}
\end{align}
The BI for these two components can be easily proven using the equation ~\eqref{G00Gij}
\begin{align}
\nabla_{\mu} G^{\mu \nu} = 0 \label{nablaGmunu} \,.
\end{align}

\subsection{Energy momentum tensor}\label{subsec:Tmunu}

To solve the EFEs given in Eq.\eqref{tEFEmp}, one must incorporate the stress-energy tensor (SET) alongside the geometric quantities from Eqs.~\eqref{Rij} and \eqref{Rmp}. In the case of a perfect fluid in thermodynamic equilibrium, which serves as the source of spacetime curvature, the SET takes the symmetric form
 \begin{align} T_{\mu\nu} = \left( \rho + \frac{P}{\tc^2} \right) U_{\mu} U_{\nu} + P g_{\mu\nu} \label{Tmunump} \,, \end{align}
 where $\rho$ represents the mass density, $P$ denotes the hydrostatic pressure, and $U^{\mu} = (\tc\,,\vec{0})$ is the four-velocity. The Einstein tensor $G_{\mu\nu}$ and the metric tensor $g_{\mu\nu}$ satisfy a key geometric identity, known as the BI \eqref{nablaGmunu} and the metric compatibility (\textit{i.e.}, $\nabla_{\mu} g^{\mu\nu} = 0$). Given the constancy of the Einstein gravitational constant $\kappa$, the BI directly leads to the local conservation of energy and momentum
 \begin{align} &\nabla_{\mu} \tensor{T}{^\mu^\nu} = 0 \quad \Rightarrow \quad \frac{\partial \rho_i}{\partial t} + 3 H \left( \rho_i + \frac{P_i}{\tc^2} \right) + 2 \rho_i H \frac{d \ln \tc}{d \ln a} = 0 \nonumber \\ &\Rightarrow d \ln \left( \rho_i \tc^2 \right) + 3 \left( 1 + \omega_i \right) d \ln a = 0 \label{BI1mp} \,, \end{align}
 where $\omega_i$ is the e.o.s parameter, defined as $\omega_i = (P_i/\tc^2)/\rho_i$.
Solving Eq.~\eqref{BI1mp} yields the following expression for the energy density
 \begin{align} &\rho_i \tc^{2} = \rho_{i0} \tc_0^2 a^{-3 (1 + \omega_i)} \label{rhomp} \,,\end{align}
where $i$ represents a perfect fluid component corresponding to either radiation or matter.

\subsection{Friedmann equations}\label{subsec:FE}

By substituting Eqs.~\eqref{Rij}, ~\eqref{Rmp}, ~\eqref{Tmunump}, and ~\eqref{rhomp} into Eq.~\eqref{tEFEmp}, the EFE for the VSL model can be expressed as
 \begin{align}
 &\frac{\dot{a}^2}{a^2} + k \frac{\tc^2}{a^2}  -\frac{ \Lambda \tc^2}{3} = \frac{8 \pi \tG}{3} \sum_i \rho_i \label{tG00mp} \,, \\
 &\frac{\dot{a}^2}{a^2} + 2 \frac{\ddot{a}}{a} +  k \frac{\tc^2}{a^2} - \Lambda \tc^2 - 2 \frac{\dot{a}^2}{a^2} \frac{d \ln \tc}{d \ln a} = -\frac{8 \pi \tG}{\tc^2} \sum_{i} P_i  \label{tG11mp} \,.
 \end{align}
These are the modified Friedmann equations due to VSL effect. From Eqs.~\eqref{tG00mp} and \eqref{tG11mp}, the equation governing the acceleration of cosmic expansion follows as
 \begin{align}
 \frac{\ddot{a}}{a} = -\frac{4\pi \tG}{3} \sum_i \left( 1 + 3 \omega_i \right) \rho_i  + \frac{\Lambda \tc^2}{3} + \frac{\dot{a}^2}{a^2} \frac{d \ln \tc}{d \ln a} \label{t3G11mG00mp} \,.
 \end{align}
By rewriting the Hubble parameter $H$ and acceleration $\ddot{a}/a$ using Eqs.~\eqref{tkappaconstmp} and \eqref{rhomp}, one obtains
 \begin{align}
 H^2 &= \left[ \frac{8 \pi \tG_0}{3} \sum_{i} \rho_{0i} a^{-3(1+\omega_i)} + \frac{ \Lambda \tc_0^2}{3} - k \frac{\tc_0^2}{a^2} \right] \frac{\tc^2}{\tc_0^2} \equiv H^{(\GR)2} \frac{\tc^2}{\tc_0^2} \label{H2me} \,, \\
 \frac{\ddot{a}}{a} &= \left[ -\frac{4\pi \tG_0}{3} \sum_i \left( 1 + 3 \omega_i \right) \rho_{0i} a^{-3(1+\omega_i)} + \frac{\Lambda \tc_0^2}{3} \right] \frac{\tc^2}{\tc_0^2} + H^2 \frac{d \ln \tc}{d \ln a} \nonumber \\
 &= \left[ \left( \frac{\ddot{a}}{a} \right)^{(\GR)} +  H^{(\GR)2} \frac{d \ln \tc}{d \ln a}  \right] \frac{\tc^2}{\tc_0^2} \label{ddotaoa} \,.
 \end{align}
These equations describe the background evolution of the FLRW universe within the VSL framework. The Hubble expansion rate in the VSL model, $H$, includes an additional factor $\tc^2/\tc_0^2$ compared to its counterpart in GR, $H^{(\GR)}$. Notably, while the present-day Hubble parameter remains unchanged between GR and VSL, its past evolution differs. A higher (or lower) speed of light in the past would have resulted in a proportionally larger (or smaller) $H$. This distinct scaling behavior has potential implications for resolving the Hubble tension.
Moreover, the acceleration of cosmic expansion in the meVSL model is modified by two key effects relative to GR. The first arises from the extra term $d \ln \tc/d \ln a$, which reflects the variation of $\tc$. The second effect corresponds to the overall scaling factor $\tc^2/\tc_0^2$, leading to a corresponding modification of the expansion rate. These deviations highlight the fundamental differences in cosmic evolution introduced by the meVSL framework.  

Although the meVSL model yields a different Hubble parameter $H(a)$ than the SMC $H^{\textsuperscript{\tiny (\GR)}}(a)$ as shown in Eq.~\eqref{H2me}, the resulting Hubble radius,
\begin{equation}
    \frac{\tilde{c}(a)}{H(a)}  = \frac{\tc_0}{H^{\textsuperscript{\tiny (\GR)}}(a)} = \frac{\tilde{c}_0}{H_0} \frac{1}{\sqrt{\sum_i \Omega_{0i} a^{-3(1+w_i)} + \Omega_\Lambda + \Omega_k a^{-2}}},
\end{equation}
remains identical to that of SMC. This is because $\tilde{c}(a)$ is constructed to preserve the causal structure of spacetime. Therefore, observables such as the luminosity distance and angular diameter distance, which are integrals over the Hubble radius as a function of redshift, will yield the same values in the meVSL framework as in SMC, as long as the redshift is defined identically~\cite{Lee:2020zts}. This implies that discrepancies between locally measured $H_0$ (e.g., from SH0ES) and values inferred from distance indicators may be reconcilable within the meVSL approach~\cite{Lee:2024nya}. However, if $\tilde{c}$ varies in a way that affects other physical constants—such as the fine-structure constant, Planck's constant, or the Chandrasekhar mass—then additional observational effects may arise~\cite{Lee:2020zts,Lee:2021xwh}. These could lead to testable differences in supernova distance moduli or in the location of the last scattering surface. We refer readers to Ref.~\cite{Lee:2020zts} for a more detailed analysis of such implications.

\section{Equation of motion of the speed of light}
\label{sec:sol}

The action given in Eq.~\eqref{SHmp} explicitly includes the Ricci scalar in Eq.~\eqref{Rmp} and the perfect fluid term, leading to the following form
\begin{align}
S &\equiv \int \Biggl[ \frac{1}{2 \tkapp} \left( \frac{6}{\tc^2} \left( \frac{\ddot{a}}{a} + \frac{\dot{a}^2}{a^2} + k \frac{\tc^2}{a^2} - \frac{\dot{a}}{a} \frac{\dot{\tc}}{ \tc} \right)  - 2 \Lambda \right) + \sum_{i=r\,,m} \rho_i (1 + \omega_i ) \tc^2 \Biggr] \sqrt{-g} dt d^3x \label{SHmppf} \,. 
\end{align}
In equation~\eqref{SHmppf}, the Lagrangian for the perfect fluid appears to depend on $\tc$ explicitly. However, from the BI equation~\eqref{rhomp} for the perfect fluid, it can be shown that this term is independent of $\tc$
\begin{align}
 \rho_i (1 + \omega_i ) \tc^2 =  \rho_{i0} (1 + \omega_i ) \tc_{0}^2 a^{-3(1 + \omega_i )}  \label{rhoi} \,.
\end{align}
Furthermore, since the Einstein gravitational constant $\tkapp$ is time-independent, as shown in equation~\eqref{tkappaconstmp}, equation~\eqref{SHmppf} can be rewritten by including only $\tc$ and $\dot{\tc}$ terms as 
\begin{align}
S &= \int \Biggl[ \frac{\tc_0^4}{16 \pi \tG_0} \left( \frac{6}{\tc^2} \left( \frac{\ddot{a}}{a} + \frac{\dot{a}^2}{a^2} + k \frac{\tc^2}{a^2} - \frac{\dot{a}}{a} \frac{\dot{\tc}}{ \tc} \right)  - 2 \Lambda \right)  \Biggr] \sqrt{-g} dt d^3x \equiv \int \frac{1}{2 \tkapp_0} R \sqrt{-g} dt d^3x \label{SHmppf2} \,.
\end{align}
From Eq.~\eqref{SHmppf2}, the Euler-Lagrange equation for \( c \) can be derived as 
\begin{align}
& \frac{d}{dt} \frac{\partial R}{\partial \dot{\tc}} - \frac{\partial R}{\partial \tc} = 0 \quad \Rightarrow \quad \frac{\ddot{a}}{a} + 3 \frac{\dot{a}^2}{a^2} = 0 \quad \Rightarrow \quad a(t) = (c_1 + c_2 t)^{1/4} \label{eomtc} \,.
\end{align}
Generally, the Euler-Lagrange equation should determine the dynamics of a given variable. However, when solving the equation for $\tc$, we find that it does not yield a dynamical equation for $\tc$ itself but instead leads to a constraint on $a(t)$. This implies that rather than being an independent dynamical degree of freedom, $\tc$ functions as a parameter that enforces a specific constraint on $a(t)$.
It means that $a(t)$ can $\tc$ be reinterpreted as an independent variable only for certain conditions on $a(t)$. This relationship suggests a strong dependence between $\tc$ and $a$, indicating that $\tc$ may not be a completely independent physical quantity. The change in $\tc$ is closely related to how we define time. If $\tc$ is always determined solely as a function of $a(t)$, it is likely because the definition of time we use has been fixed in a specific way. Therefore, if we can redefine time, we may be able to treat $\tc$ as an independent variable. On the other hand, if $\tc$ always depends on $a(t)$ regardless of the choice of the time variable, then $\tc$ is likely to be fundamentally a dependent variable.

\begin{figure}
	\begin{center}
	\includegraphics[width=0.9\textwidth]{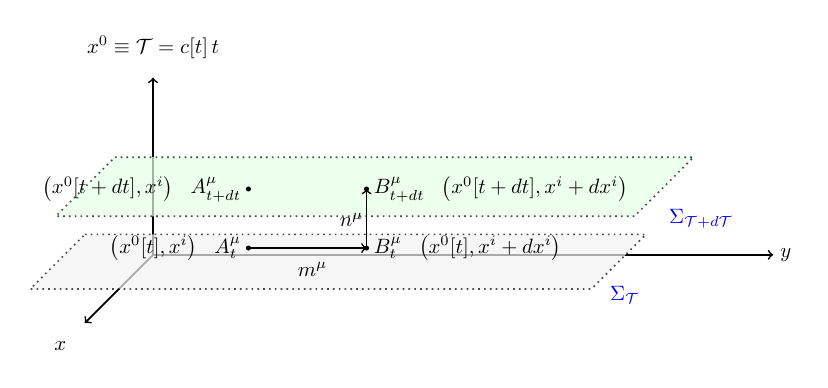} 
	\caption{This illustrates the foliation of spacetime within the RW metric framework. The hypersurfaces $\Sigma$ represent constant values of the temporal coordinate $\mathcal{T}$, corresponding to different moments in time. The lapse function governs the separation between these hypersurfaces, ensuring that the normal vector $\bar{n} = n^{\mu} e_{\mu}$ remains orthogonal to them.}
	\label{Fig2}
	\end{center}
\end{figure}

In meVSL model, the lapse function accounts for the variation of $\tc$. We can define the lapse function $N \equiv \tc/c$ as 
\begin{align}
x^{0}(t+dt) - x^{0}(t) &=  \left( c[t+dt] \right) \left( t+dt \right) - c(t) t \overset{\mathcal{O}(1)}{\approx} c[t] dt + dc[t] t \nonumber \\
	&= \left( 1 + \frac{d \ln c}{d \ln t} \right) c[t] dt \equiv \tilde{c}[t] dt \equiv N[t] c[t] dt \label{NmeVSL} \,.
\end{align}
The ability to represent the variation of $\tc$ through the lapse function implies that the change in $\tc$ is not merely a physical dynamical process but depends on the definition of the time variable we choose. Therefore, $\tc$ is more likely to be determined by the choice of coordinates (lapse function) rather than being an independent variable. In other words, it is necessary to reconsider whether $\tc$ should be treated as an independent dynamical degree of freedom or interpreted as a coordinate effect through the lapse function.

In the meVSL model, the lapse function represents the flow of time on each hypersurface. If there is no mechanism that determines the lapse function, then the Euler-Lagrange equation for $\tc$ does not describe a dynamical evolution but only imposes a constraint equation.  The fact that $\tc$ can be expressed through the lapse function implies that it is not an independent dynamical degree of freedom but rather a quantity determined by the choice of coordinates. Therefore, without an additional mechanism to fix the lapse function,  $\tc$ cannot be treated as a truly independent variable. 

The lapse function $N(t)$ determines the rate at which proper time flows relative to the coordinate time in a given foliation of spacetime. In the ADM decomposition, it reflects the arbitrariness of time slicing and encodes how time advances from one spatial hypersurface to another. Physically, it sets the clock rate experienced by comoving observers, meaning that a varying lapse corresponds to a nontrivial temporal gauge. In the context of the meVSL framework, this allows us to interpret the time dependence of the effective speed of light $\tc(t)$ as arising from a choice of lapse function, rather than from an independent field dynamics. This interpretation preserves the general covariance of GR and reframes the variation of $\tc$ as a coordinate effect rather than a sign of new physics.

\section{Conclusion and Summary}\label{sec:Conc}

In this work, we have examined the meaning of a Varying Speed of Light within the Robertson-Walker metric. We have demonstrated that the variation of $\tc$ in this context does not necessarily imply a fundamental physical change in the speed of light but reflects a coordinate effect associated with the choice of cosmic time.

A key observation is that the cosmic time in the RW metric is identical to the proper time of comoving observers, as ensured by the Weyl postulate. Under this setting, a varying c corresponds to a change in the clock rate across hypersurfaces rather than an intrinsic physical evolution. This interpretation is reinforced by deriving the Euler-Lagrange equations from an action that includes $\tc$. The resulting equations do not describe the dynamics of $c$ itself but instead impose a constraint on the evolution of the scale factor $a(t)$. This result implies that $\tc$ is not an independent degree of freedom but depends on the cosmic time-lapse function.

This reinterpretation of VSL challenges conventional assumptions about its role in cosmology. Many VSL models assume that a changing $\tc$ represents a fundamental modification of spacetime physics, often introducing additional assumptions or mechanisms to justify such variation. However, our analysis suggests that these changes can be understood as a result of the coordinate choice rather than requiring new physics. In this view, the apparent evolution of $\tc$ reflects how the universe's temporal structure is parameterized, rather than an actual modification of light propagation.
In conclusion, our findings suggest that VSL is a coordinate-dependent feature of cosmic time rather than a fundamental modification of physical laws. This insight offers a new perspective on VSL models and their implications for cosmology, particularly in the context of cosmic expansion and observational tensions. Further investigation is needed to explore whether this framework can provide a more consistent and natural resolution to key issues in modern cosmology.

This approach may also shed light on current observational tensions such as the Hubble tension or time dilation anomalies in Type Ia supernovae, offering a reinterpretation without invoking new dynamical degrees of freedom. Further investigation is needed to determine whether this framework can consistently reproduce cosmological observables, such as luminosity distance–redshift relations, and whether it offers viable alternatives to the standard model cosmology. As a gauge choice, we may arbitrarily set the form of the varying speed of light as $\tc = \tc_0 a^{b/4}$. By comparing this parametrization with cosmological observations such as time dilation measurements from Type Ia supernovae and other probes, we can test whether the value of $b$ deviates from zero. If observational data favor $b \neq 0$, it would indicate that the commonly adopted gauge choice $b = 0$—which corresponds to the standard cosmological model—may not accurately describe the temporal structure of our universe. Conversely, if $b = 0$ remains favored, it would reinforce the consistency of the standard gauge with observations.


\acknowledgments{This research was funded by the National Research Foundation of Korea (NRF), funded by the Ministry of Education (Grant No. NRF-RS202300243411).}

\end{document}